\documentclass{mn2e} 
\usepackage{color}
\usepackage{dcolumn}
\usepackage{bm}
\usepackage{epsfig}

\def\reference{\par\noindent\hangindent\parindent}

 at 13 truept

\begin{document}

\twocolumn

\title[Two-phase galaxy evolution]{Two-phase galaxy evolution: the cosmic star-formation histories of spheroids and discs}

\author[Driver et al.] 
{S.P.~Driver$^{1,2}$\thanks{SUPA, Scottish Universities Physics Alliance}, 
A.S.G.~Robotham$^{1,2}$, 
J.~Bland-Hawthorn$^{3}$
M.~Brown$^4$,
A.~Hopkins$^{5}$, 
\newauthor 
J.~Liske$^{6}$, 
S.~Phillipps$^{7}$, 
S.~Wilkins$^8$\\
$^1$ International Centre for Radio Astronomy Research (ICRAR), University of Western Australia, Crawley, WA 6009, Australia \\
$^2$ School of Physics \& Astronomy, University of St Andrews, North Haugh, St Andrews, KY16 9SS, UK; SUPA \\
$^3$Sydney Institute for Astronomy, School of Physics, University of Sydney, NSW 2006, Australia\\
$^4$School of Physics, Monash University, Clayton, Victoria 3800, Australia\\
$^5$ Australian Astronomical Observatory, PO Box 296, Epping, NSW 1710, Australia\\
$^6$ European Southern Observatory, Karl-Schwarzschild-Str.~2, 85748 Garching, Germany\\
$^7$ HH Wills Physics Laboratory, University of Bristol, Tyndall Avenue, Bristol, BS8 1TL, UK\\
$^8$ School of Physics and Astronomy, Oxford University, Keeble Road, Oxford, UK \\
}
\pubyear{2012} \volume{000}
\pagerange{\pageref{firstpage}--\pageref{lastpage}}

\maketitle
\label{firstpage}

\begin{abstract}
From two very simple axioms: (1) that AGN activity traces spheroid
formation, and (2) that the cosmic star-formation history is dominated
by spheroid formation at high redshift, we derive simple expressions
for the star-formation histories of spheroids and discs, and their
implied metal enrichment histories.

Adopting a Baldry-Glazebrook initial mass function we use these
relations and apply {\sc pegase.2} to predict the $z=0$ cosmic
spectral energy distributions (CSEDs) of spheroids and discs. The
model predictions compare favourably to the dust-corrected CSED
recently reported by the Galaxy And Mass Assembly (GAMA) team from the
FUV through to the $K$ band. The model also provides a reasonable fit
to the total stellar mass contained within spheroid and disc
structures as recently reported by the Millennium Galaxy Catalogue
team. Three interesting inferences can be made following our axioms:
(1) there is a transition redshift at $z \approx 1.7$ at which point
the Universe switches from what we refer to as ``hot mode evolution"
(i.e., spheroid formation/growth via mergers and/or collapse) to what
we term ``cold mode evolution" (i.e., disc formation/growth via gas
infall and minor mergers); (2) there is little or no need for any
pre-enrichment prior to the main phase of star-formation; (3) in the
present Universe mass-loss is fairly evenly balanced with
star-formation holding the integrated stellar mass density close to a
constant value. 

The model provides a simple prediction of the energy output from
    spheroid and disc projenitors, the build-up of spheroid and disc
    mass, and the mean metallicity enrichment of the Universe.
\end{abstract}


\section{Introduction}
The vast majority of galaxies can be adequately described as
consisting of a compact smooth spheroidal component containing a
predominantly pressure-supported old [$\alpha$/Fe]-enhanced stellar
population, and/or an extended flattened star-forming disc component
containing intermediate and young stars with a wide range in
metallicities, having smooth rotation and embedded in an extensive gaseous
cold gas disc. Exceptions exist, most notably the dwarf populations
which, while dominant in terms of number density, actually contribute
only a modest amount to the baryon budget at the present time ($<16$
per cent; Driver 1999; Geller et al.~2012). This dichotomy of galaxies
into spheroids and discs has been known for over a hundred years
stretching back to even before the confirmation that galaxies are
external systems (e.g., Hubble~1926, 1936; Zwicky~1957 and references
therein). To some extent this dichotomy has been recently
``rediscovered'', through the statistical studies of large populations
as a galaxy bimodality (Strateva et al.~2001; Baldry et al.~2004).
Driver et al.~(2006) argued that this bimodality is better
interpreted in terms of the earlier bulge-disc dichotomy and advocated
routine structural decomposition as vital (e.g., Allen et al.~2006;
Simard et al.~2011; Lackner \& Gunn 2012) to directly trace the
independent evolutionary histories of the spheroidal and disc
components.

Numerical models of galaxy formation struggle to produce realistic
galaxy systems with a tendency to form overly cuspy cores and
difficulty in maintaining extended disc structures with a high axis
ratio (White \& Navarro~1993; Navarro \& Steinmetz~2000; Abadi et
al.~2003; House et al.~2011). Both problems are likely to be connected
to the different fundamental properties of the dark matter and the
baryons, and in particular their ability to experience pressure and
their ability to dissipate energy. In the core regions the
gravitational coupling of the baryons with the dark-matter may allow
it to exhibit a pseudo-pressure, whereas in the outer-regions the
ability of the baryons to dissipate energy on a timescale which is
faster than the free-fall timescale may allow for the formation
of a thin rotating baryonic disk. This picture while simple to
articulate has proven extremely hard to simulate, with the need to
partition and redistribute the angular momentum in a quite specific
manner to result in galaxies with realistic appearances. In particular
merger events are extremely disruptive to this process, imparting both
energy and angular momentum to the baryonic disc, which is easily
disrupted or 'plumped up' (see Barnes \& Hernquist 1992 for extensive
discussion and early references on this topic, also Hopkins et
al.~2009 for updated simulations on the survivability of discs during
merger events). In general the greater the merger-rate the more
bulge-dominated the final galaxy population appears.

More contemporary hydrodynamical simulations (e.g., Governato et
al.~2010; Agertz, Romain \& Moore~2011; Scannapieco et al.~2011;
Domenech-Moral et al.~2012) are now starting to show significant
success at producing realistic ``looking'' bulge-disc systems by
incorporating a greater level of cold gas infall than previously
assumed, as argued earlier by Keres et al.~(2005) and Dekel et
al.~(2009). These focused hydrodynamical studies, however, are
inevitably extracted from numerical simulations with particularly
quiescent merger histories, suggesting such systems should be the
exception rather than the norm. Hence, while numerical simulations of
the development of the dark matter haloes find a continual process of
halo merging, it appears that the baryons and what we identify as
galaxies (baryonic condensates), might not develop in the same
way. Martig \& Bournaud (2009) argued that feedback from low and
intermediate mass stars can contribute significantly to the
redistribution of mass from the bulge to the disc through extensive
(or even excessive) mass-loss. This baryonic outflow could help
alleviate the problem of excessive bulge-formation by allowing some
fraction of the collapsed stellar mass (up to $\sim 50$ per cent),
to return to the halo and contribute to the later growth of a disc ---
thereby coupling bulge and disc growth. However this mechanism also
relies on a fairly quiescent merger history during later times, and
does not easily explain the broad diversity of bulge-disc ratios
seen. As an aside, simulations have also demonstrated that baryonic
outflows from the core regions can help provide a plausible
explanation to the core-cusp problem (Governato et al.~2012; Zolotov
et al.~2012).  Clearly feedback and infall are both crucial processes,
whose motivation is as much driven by the requirement to produce
realistic looking images as by fundamental physics, and which are both
more effective if the merger rate is either low or at least confined
to earlier epochs.

As argued in the opening paragraph, we advocate a more heuristic
approach where we put aside the issue of dark-matter assembly and
start by asking whether the dichotomy of galaxy structure is best
explained by two distinct formation mechanisms. Following the earlier
discussion and lessons learnt from the simulations, the obvious two
mechanisms can loosely be termed as a hot and cold mode. In the hot
mode spheroids are formed early and rapidly via dynamically hot
(turbulent) processes (collapse, fragmentation, and merging). In the
cold mode discs are formed more slowly, from an extended quiescent
phase of cold gas infall regulated by internal feedback (i.e.,
supernova). This basic concept is of course not new (e.g., Larson
1976; Tinsley \& Larson 1978) but has laid dormant for sometime
overshadowed by the dominance of merger-driven evolution. However the
revival is also being championed via a series of semi-analytic studies
by Cook et al.~(2009; 2010a; 2010b, see also Dekel et al.~2009)
inspired by behaviour seen in numerical simulations in which an
initial rapid hot merger phase is typically followed by a more
quiescent phase of accretion (see also L'Huillier, Combes \& Semelin
2012).

The two-phase model is both obvious (given the bulge-disc nature of
galaxies) and controversial, as it marginalises the merger rate
required for dark matter assembly to earlier epochs than simulations
typically suggest. This low merger rate is arguably corroborated by
the local studies of dynamically close pairs (in particular see Patton
et al.~2002; De Propris et al.~2005, 2007, 2010) --- although a
correct derivation of the merger rates requires a robust understanding
of the merger timescales, which are currently poorly
constrained. Perhaps more compelling, however, is the result that only
40\% of the present day stellar mass resides in spheroidal systems
(Driver et al.~2008; Gadotti~2009; Tasca \& White~2011). A key
inference is then: {\it If discs are destroyed/thickened during
  mergers, yet the majority of stellar mass resides in discs, the
  dominant formation mechanism cannot be merger-driven, but presumably
  the more quiescent process of cold gas accretion.} This statement
becomes more profound when one realises that the stellar mass in discs
today only measures that unaffected by mergers, and that some of the
stars currently in spheroidal systems may have originally formed
within discs via cold accretion prior to a merger event. In some bulge
formation scenarios, star-formation via merging is dispensed with
altogether and replaced by the migration of massive star-formation
clumps formed within deeply turbulent discs (e.g., Elmegreen, Bournaud
\& Elmegreen 2008). This potentially relegates the stellar-mass
build-up driven by mergers to be in the 0---40 per cent range by
mass. Clearly mergers do occur at all redshifts and similarly discs
may form, be disrupted, and reform at any redshift. Recently
L'Huillier, Combes \& Semelin (2012) reported that 77 per cent of a
galaxy's mass is formed via gas accretion and 23 per cent via direct
merging from simulations. Other empirical studies also seem to suggest
that the bulk ($\sim 70$ per cent) of the stellar mass is mostly
assembled by $z \sim 1$, again marginalising the role of late time
major mergers (e.g., Bundy et al.~2004; Brown et al.~2007, 2008).

Focused studies of nearby galaxies are also unveiling significant
levels of gas accretion in some nearby systems (Sancisi et al.~2008) and
studies of the very rapid evolution of galaxy sizes have argued
(e.g., Graham et al.~2011) that the compact elliptical systems seen at
intermediate redshift ($1.4 < z < 2.5$) by Daddi et al.~(2005) and
Trujillo et al.~(2006) (see also Bruce et al.~2012) might represent
the naked bulges of present day spiral systems.

In essence the two-phase model is an attempt to highlight,
conceptually, the possibility of a distinct change in the primary galaxy
formation mechanism occurring at some transition redshift from an era
where the {\it dominant} mode is major mergers leading to spheroid
formation, to an era where the {\it dominant} mode is accretion
leading to disc formation.

At early cosmic epochs we see a prevalence of distinct
phenomena, in particular highly asymmetrical morphology in
massive/luminous systems (Driver et al.~1998; Conselice, Blackburn \&
Papovich.~2005; Ravindranath et al.~2006) and significantly increased
AGN activity (Fan et al.~2001,2003; Croom et al.~2004; Richards et
al.~2006). AGN activity is directly linked to the formation and growth
of the associated super-massive black holes (SMBH; Hopkins et
al.~2008a) which in turn is linked to spheroid formation via the well
established SMBH-bulge relations (see, for example, the review by Ferrarese
\& Ford 2005 or the recent near-IR SMBH-bulge luminosity relation by
Vika et al.~2012). Recent studies also argue, from more direct
empirical evidence, that AGN activity is almost always coincident with
massive star-formation and that the two-processes do indeed appear to
occur hand-in-hand (e.g., Rafferty et al.~2011). This AGN-SMBH-bulge
connection therefore implies a clear timescale for the formation of
the spheroid systems (see also Pereira \& Miranda~2011 for a similar
argument, albeit applied in the opposite direction).

In Section 2 we describe the $z=0$ empirical data describing the
cosmic spectral energy distribution (CSED) of spheroids and discs as
recently reported by the Galaxy And Mass Assembly team (Driver et
al.~2011, 2012). In Section 3 we take the above arguments to their
natural conclusion and use the AGN-SMBH-Bulge connection to define the
independent star-formation history of spheroids and assign the
residual star-formation, implied by the cosmic star-formation history,
to describe that of the discs. In Section 4 we use our star-formation
histories to produce predictions of the CSED of spheroids and discs,
and in Section 5 compare the predictions to the data.

Throughout we use $H_0$=70 h km s$^{-1}$ Mpc$^{-3}$ and adopt
$\Omega_M=0.27$ and $\Omega_{\Lambda}=0.73$ (Komatsu et al.~2011).

\section{The $z=0$ cosmic spectral energy distribution}
In Driver et al.~(2012) we reported the empirical measurement of the
cosmic spectral energy distribution (CSED) in the nearby Universe,
corrected for dust attenuation, and spanning the wavelength range from
0.1 to 2.1 $\mu$m, i.e., the regime over which direct stellar-light
dominates. These data were derived from the combination of the GAMA
spectroscopic survey, currently underway on the Anglo-Australian
Telescope (Driver et al.~2011), coupled with reprocessed and aperture
matched data from GALEX, SDSS, and UKIRT LAS (see Seibert et al., in
prep. and Hill et al.~2010). Driver et al.~(2012) also provided the
CSED subdivided according to spheroid-dominated and disc-dominated
systems. The division into spheroid and disc dominated was achieved
via visual classification, as neither a simple colour nor S\'ersic
index division appears to cleanly separate the two populations (see
also Kelvin et al.,~2012, figure~20).

The sample originated from a common volume of $2.8 \times 10^5$
(Mpc/h)$^{3}$ over the redshift range $0.013 < z < 0.1$.  Although the
GAMA survey currently contains about 180,000 galaxies with known
redshifts, the adopted redshift range significantly reduces the sample
size to around 10,000. It also simplifies and removes any luminosity
bias arising from large scale clustering as the sample is
pseudo-volume limited around the $L^*$ region --- i.e., those galaxies
which dominate the luminosity density measurements.

As the GAMA survey lies entirely within the Sloan Digital Sky Survey,
the GAMA CSED was renormalised to the full SDSS survey area. This
reduces the cosmic/sample variance from around 15 per cent to 5 per
cent (using the formula for estimating cosmic variance given by eqn.~4
of Driver \& Robotham 2010).

\subsection{The CSED of spheroids and discs}
The final CSED values we use here are the spheroid-dominated and {\it
  attenuation corrected} disc-dominated values taken directly from
Table~7 of Driver et al.~(2012). 

As dust attenuation is such a crucial issue it is worth mentioning the
genesis of the corrections used by the GAMA team. First, the dust
correction is only applied to the disc-dominated data and the spheroid
population is assumed dust free (e.g., Rowlands et al.~2012).  Second,
the corrections are based on the radiative transfer models of Tuffs et
al.~(2002) and Popescu et al.~(2011) which have been fine-tuned to the
multiwavelength (FUV--far-IR) data of NGC891, and incorporate three
distinct dust components; an extended low opacity disc, a compact high
opacity disc and dust clumps. This fiducial model is calibrated to the
galaxy population at large by modifying the $B$-band central face on
opacity until the predicted variation of flux with inclination matches
the trend of $M^*$ with inclination seen in the Millennium Galaxy
Catalogue data (Driver et al.~2007). This calibrated model was then
used to derive the combined face-on and inclination dependent
correction for a population of galaxies averaged over all viewing
angles and over a wavelength range of $0.1-2.1\,\mu$m (Driver et
al.~2008). This photon escape fraction (varying from 24 per cent in
the FUV to 89 per cent in the $K$-band) was then used to correct the
CSED of disc-dominated systems to the intrinsic CSED, which we use
here.

The CSED of spheroid-dominated and disc-dominated galaxies, however,
is not precisely what we require, as some proportion of the CSED flux
in the disc-dominated class may be coming from the central bulges.
Likewise, some proportion of the flux in the spheroid-dominated class
may be due to faint discs. In order to assess how much of a problem
this might be, we can compare the ratio of the $K$-band luminosity
densities of the spheroid-dominated to non-spheroid dominated samples,
to the ratio of the stellar mass densities of bulge+elliptical systems
to disc systems from Driver et al.~(2007)\footnote{The Driver et
  al.~(2007) study is based on bulge-disc decompositions of the
  Millennium Galaxy Catalogue data (Liske et al.~2003; Driver et
  al.~2005) described in full in Allen et al.~(2006)}. This test
assumes that the $K$-band luminosity is a suitable single-band proxy
for stellar mass. We find reasonable agreement (within $12$ per cent),
suggesting that a comparable amount of flux needs to be redistributed
in either direction. In detail, the $K$-band luminosity densities are
1.2 and 2.2 ($\times 10^{34}$ h W Mpc$^{-3}$) for the
spheroid-dominated and non-spheroid-dominated populations respectively
(taken from Table~7 of Driver et al.~2012). Meanwhile the stellar mass
densities for spheroids (bulge+ellipticals) and discs are 2.9
and 4.7 ($\times 10^8$h M$_{\odot}$ Mpc$^{-3}$) respectively (taken
from Table~1 of Driver et al.~2007). This gives agreement to $\sim 10$
per cent and suggests that for the moment we can adopt the following
approximation:

~

elliptical+bulge CSED $\approx$ Spheroid-dominated CSED

~

disc CSED $\approx$ non-Spheroid dominated CSED

~

In due course all galaxies at $z<0.1$ in the GAMA survey will be
decomposed into bulge and disc components to enable a direct
derivation of the true spheroid and disc CSEDs.

\section{The star-formation history of spheroids and discs}
The local CSED should be a predictable quantity if the cosmic
star-formation history (CSFH) is known, the initial mass function
(IMF) is universal and known, and a plausible stellar evolution code
applied. Of course this is not quite so simple and in particular the
metallicity enrichment adopted will significantly modify the
predicted CSED shape. Upcoming papers will explore these issues in
more detail, but here we wish to construct a basic first-look model
and focus on the viability of the hypothesis that galaxy formation
progressed in two fairly distinct phases: rapid spheroid formation
followed by more quiescent disc growth.

In order to construct a model of the present day spheroid and disc
CSEDs we need not just the CSFH, but the CSFH sub-divided into
spheroids and discs. These CSED predictions can then be compared to
the data from Section 2. 

The existence of various super-massive black-hole bulge relations
(e.g., Ferrarese \& Ford 2005), provides the obvious smoking gun, as
it couples SMBH growth to bulge growth. This is because SMBHs are
believed to grow via mergers, resulting in an active-galactic nucleus
phase (Hopkins et al.~2006). The growth of spheroids is therefore,
arguably, mirrored via the more readily observable AGN activity
history. This logical connection, from a correlation to causality, is
the key assumption underpinning our model and forms the first of our
two axioms.  In the recent study by Richards et al.~(2006), the
integrated AGN activity versus redshift was reported and, ignoring any
significant lag (in either spheroid formation or AGN activity), can be
used as a proxy in shape for the spheroid cosmic star-formation
history. 

The amplitude of the spheroid SFH can be set from comparison of the
AGN activity shape to the global CSFH. For our second axiom we elect
to maximise the spheroid CSFH by setting the amplitude as high as
possible without exceeding the global CSFH (i.e., a maximal spheroid
formation scenario).  Conceptually then, the heart of the two-phase
model can be defined empirically from two axioms:

\begin{figure*}

\centerline{\psfig{file=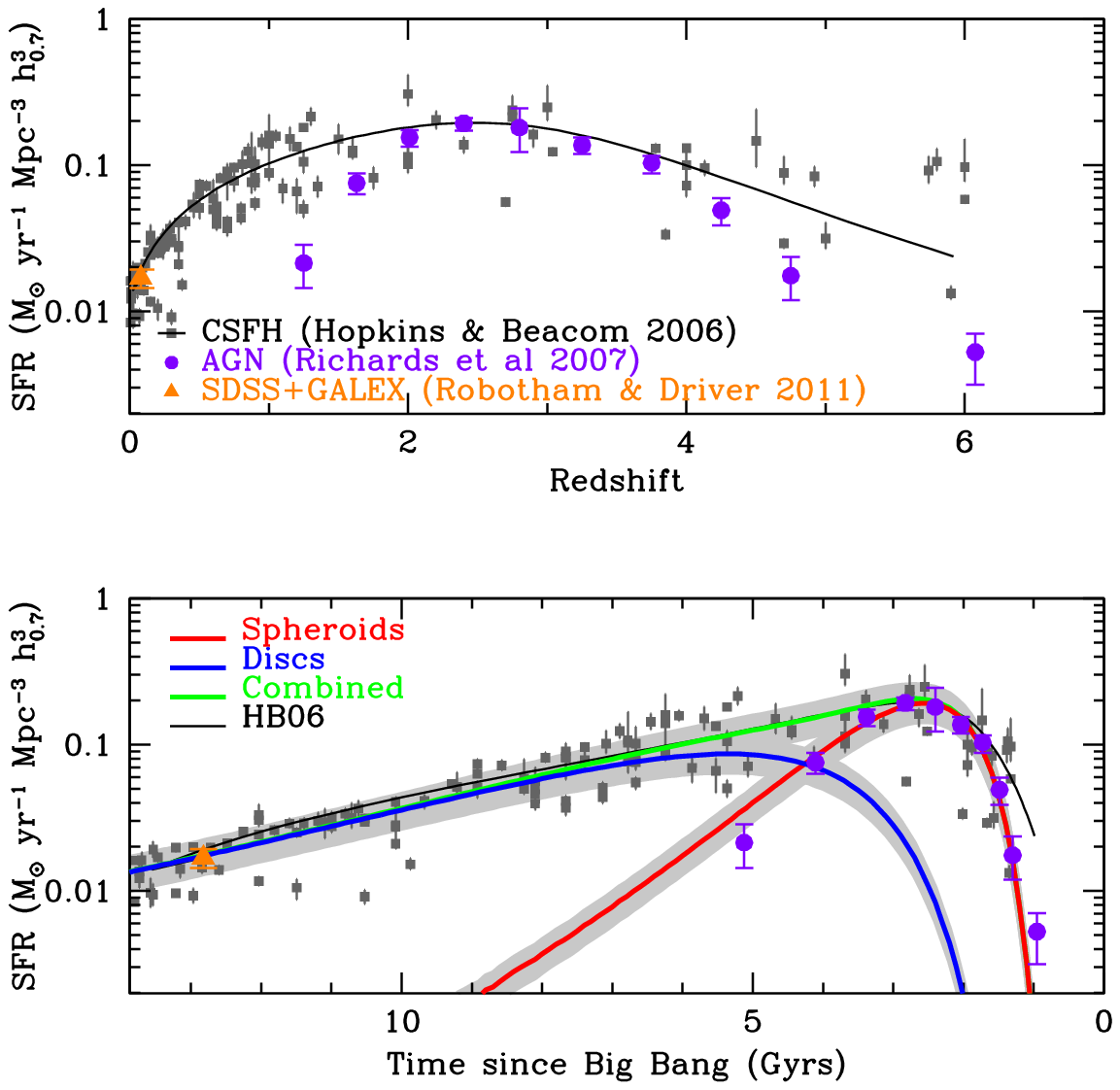,width=\textwidth}}


\caption{\label{fig:csfh} ({\it upper panel}) the cosmic
  star-formation history from Hopkins \& Beacom~(2006), Table~2 (black
  line) and the actual data (grey points) calibrated to a
  modified-Salpeter IMF (see Table~\ref{tab:mult}). Overlain are the
  QSO luminosity density data from Richards et al.~(2006), Figure
  20. The QSO luminosity density is scaled until the peak aligns with
  the peak of the CSFH. ({\it lower panel}) the same data but now
  shown with the ordinate in units of age.  Overlain are our
  parametric fits to these data which provide our inferred CSFHs for
  the Spheroid and Disc systems. On both panels we also show the SFR
  derived from the SDSS/GALEX FUV luminosity function given in
  Robotham \& Driver (2006) converted to a modified-Salpeter A IMF
  (see Table~\ref{tab:mult} or Hopkins \& Beacom 2006).}
\end{figure*}

~

{\bf 1) AGN activity traces spheroid growth}

~

{\bf 2) Spheroid formation dominates at high-z}

~

As the above two axioms are already constrained by empirical data,
this provides a zero-parameter starting point for the two-phase model
--- bypassing the need for any initial conditions or detailed
numerical simulations. Fig.~\ref{fig:csfh} (upper) shows the fit (solid
curve) to the cosmic star-formation history data (grey data points)
taken from Hopkins \& Beacom~(2006; see their figure~2a and table~1
column 2). This adopts the parametric form defined by Cole et
al.~(2001) and where the UV data has been calibrated to a
modified-Salpeter~(1955) IMF. The data describing the AGN luminosity
density are taken from Richards et al.~(2006) and rescaled such that
the peak of the AGN luminosity density lies on the CSFH curve
(requiring an arbitrary multiplication by a factor of $3.51 \times
{10^6}$M$_{\odot}$/yr$^{-1}$L$^i_{\odot}$). Immediately noticeable is
the apparent discrepancy/uncertainty at very high
redshift. Particularly as the axioms above require that at the very
highest redshift the CSFH and AGN activity curves should have the same
form. To some extent the evident discrepancy simply reflects data
uncertainty, as the dust corrections on the CSFH at high-z are poorly
constrained with significant ongoing debate as to the true shape of
the CSFH at the highest redshifts. For example, measurements of the
star-formation history based on gamma-ray bursts (Yuksel et al.~2008;
Kistler et al.~2009) often find higher star-formation rates. Likewise
the incidence of dust-obscured AGN is an equally hotly debated topic
(e.g., Polletta et al.~2006; Triester et al.~2010). Most
    recently Behroozi, Wechsler \& Conroy (2012) argue that the
    compendium by Hopkins \& Beacom (2006) potentially leads to an
    over-estimate of the cosmic star-formation history at very high
    redshifts as the pre-2006 UV luminosity densities may have been
    over-estimated and find a modified CSFH which agrees very well
    with the high-z AGN data (see their figure 2).

When presented as a function of time (Fig.~\ref{fig:csfh}, lower), it
is clear that this discrepancy is not actually that significant, and
the accumulated stellar mass able to be formed during this high-z
interval is small compared to subsequent mass growth. We can now
trivially fit the modified-Richards data to derive the star-formation
history of spheroids. This fit can then be subtract from the global
star-formation history and in turn fit to recover the implied disc
formation history. The resulting expressions are given below and
represent the star-formation rate ($\dot{\rho}$) versus time in Gyrs
($t_{\rm Gyrs}$) since the Big Bang for the spheroidal ($S$) and disc
($D$) populations:
\begin{eqnarray}
\dot{\rho}_{S}=\xi 1.03 \times 10^{-5} h_{0.7}^{3} (\frac{21.86}{t_{\rm Gyrs} h_{0.7}})^{8.57}\exp(-\frac{21.86}{t_{\rm Gyrs} h_{0.7}}) \\
\dot{\rho}_{D}=\xi 1.80 \times 10^{-3} h_{0.7}^{3}(\frac{29.39}{t_{\rm Gyrs} h_{0.7}})^{5.50}\exp(-\frac{29.39}{t_{\rm Gyrs} h_{0.7}})
\end{eqnarray}
where $\xi$ is the IMF multiplier as given in
Table~\ref{tab:mult}. The guideline error at any particular time
should be taken as $\sim \pm 25$\% based on the scatter of the
original data shown on Fig.~\ref{fig:csfh} (lower) i.e., $\sim 70$\%of
the grey data points lie on the grey shaded regions. At this point it
is also worth highlighting that while the AGN data (mauve points)
place a strong constrain on the shape of the spheroid star-formation
history the normalisation is very uncertain and based on the
relatively limited high-z cosmic star-formation history data and as we
shall see in Section.~4 will require a modest downward correction of
25\% to match the observed CSED (i.e., within the grey shaded error
bounds).

~

As an alternative for the spheroid population, one could instead use
the total CSFH for $t<3$Gyrs combined with the AGN luminosity density
data for $t>3$Gyrs, which is given by:
\begin{equation}
\dot{\rho}_{S2}=\xi 2.72 \times 10^{-4} h_{0.7}^{3} (\frac{16.82}{t_{\rm Gyrs} h_{0.7}})^{6.97}\exp(-\frac{16.82}{t_{\rm Gyrs} h_{0.7}})
\end{equation}
As stated these CSFHs are calibrated, via the UV data, to the
modified-Salpeter IMF used by Hopkins \& Beacom (2006) which was first
laid down as Sal A by Baldry \& Glazebrook (2003). To convert to other
IMFs one needs to {\it multiply} by the factor ($\xi$) shown in
Table~\ref{tab:mult}.
\begin{table}
\caption{CSFH multiplication factors ($\xi$) for various
  IMFs. \label{tab:mult}}
\begin{tabular}{ll} \hline
IMF & Multiplier \\ 
    &  $\xi$ \\ \hline
Salpeter~(1955) & $\times 1.3$ \\
modified-Salpeter (Hopkins \& Beacom 2006) & $\times 1.0$ \\
Baldry \& Glazebrook~(2003) & $\times 0.7$ \\
Kroupa~(1993) & $\times 1.7$ \\
Kroupa~(2001) & $\times 0.85$ \\ 
Chabrier (2003) & $\times 0.85$ \\ \hline
\end{tabular}
\end{table}
The simple expressions above are shown in Fig.~\ref{fig:csfh} (lower)
in red (spheroid), blue (disc), and green (spheroid + disc) and
provide a good fit to the data given the accuracy to which the data
are known. These equations now provide a blueprint for the formation
of the present day spheroids and discs over the full age of the
Universe, leading to a clear prediction of the stellar energy output
and stellar mass growth. 

Of particular interest should be the transition point around 4.2\,Gyrs
($z \approx 1.6$) from which point star-formation resulting in disc
growth dominates over star-formation resulting in spheroid
growth. This suggests a key epoch at which the Universe switches from
merger dominated evolution to accretion dominated evolution and meshes
very well with the evident change in the morphological appearance of
galaxies from highly disturbed to more ordered systems at $z \sim 1.5$
(see Driver et al.~1998; and also van den Bergh 2002).

\section{Constructing the model}
With the CSFH of spheroids and discs defined, we now build our
empirical two-phase model adopting ``vanilla'' choices at every
opportunity. In summary the key inputs and assumptions to the model are:

~

\noindent
1) The star-formation history shown in Fig.~1 as discussed in section~3.

\noindent
2) The adoption of a Universal IMF, in this case Baldry \& Glazebrook~(2003, henceforth BG03).

\noindent
3) The adoption of {\sc pegase.2} (Fioc \& Rocca-Volmerange~1997; 1999) to
model the spectral output of the evolving stellar population (using default options throughout).

\noindent
4) The assumption that the gas-phase metallicity increases
linearly with star-formation from $Z=0.0$ to $Z=0.030$ for spheroids
and to $Z=0.010$ for discs, with no time lag (i.e., instantaneous
enrichment).

~

\begin{figure}

\centerline{\psfig{file=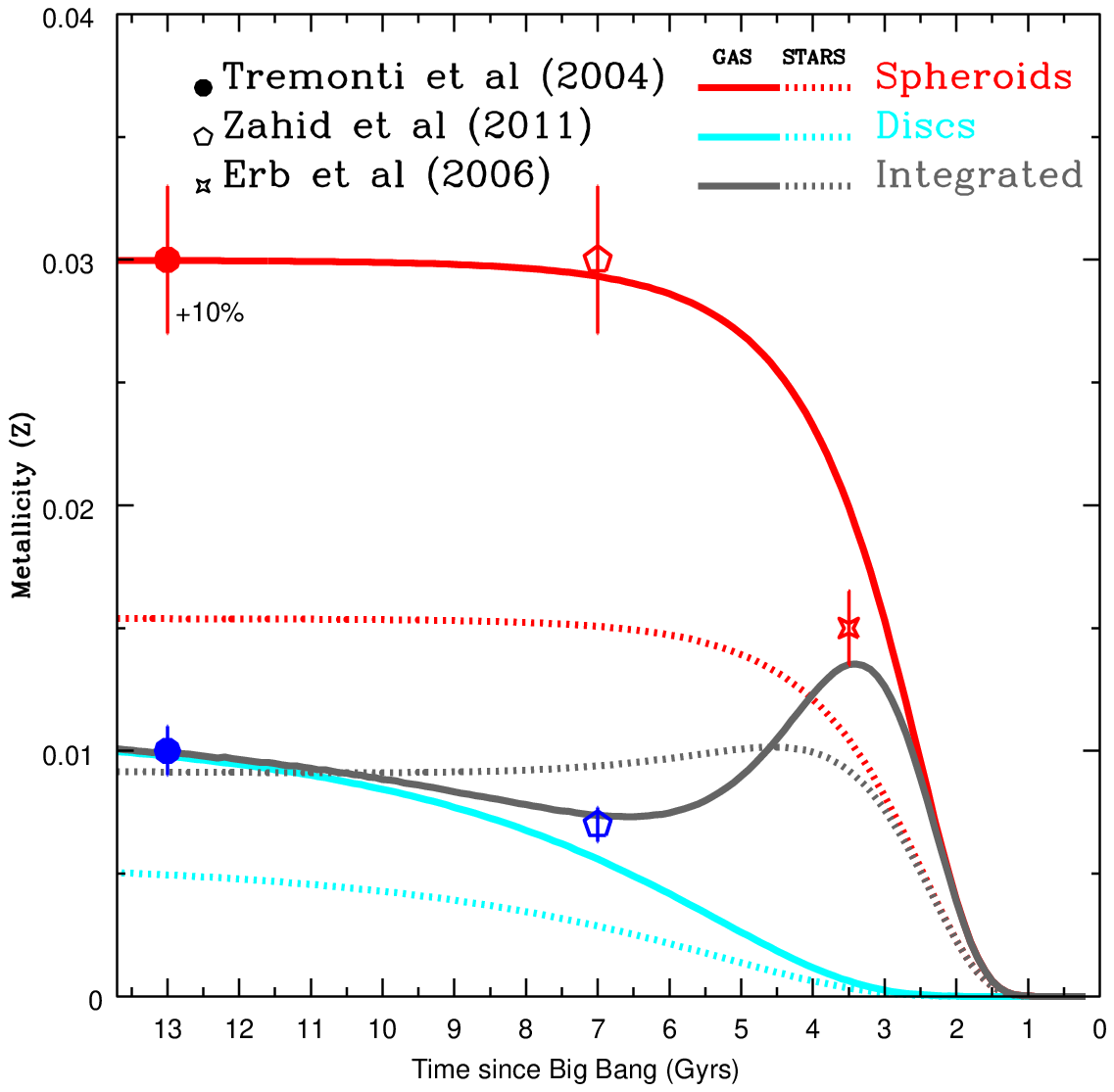,width=\columnwidth}}

\caption{\label{fig:metal} The adopted (gas-phase) metallicity for
  spheroids (red solid line) and discs (cyan solid line) as a function
  of time. Solid data points show the accepted mean $z=0$ values taken
  from Tremonti et al.~(2004). Also shown are approximate data values
  from Erb et al.,~(2006) and Zahid et al.,~(2011). Note that all data
  have an arbitrary error of $\pm 10$ \%. The gas-phase metallicity
  assumes no lag between star-formation and enrichment. The dotted
  lines show the implied metallicity of the stellar populations for
  the spheroids and discs. The grey line shows the mean by mass of the
  integrated stellar metallicity and the instantaneous integrated
  gas-phase metallicity.}
\end{figure}

\begin{figure*}[h]

\centerline{\psfig{file=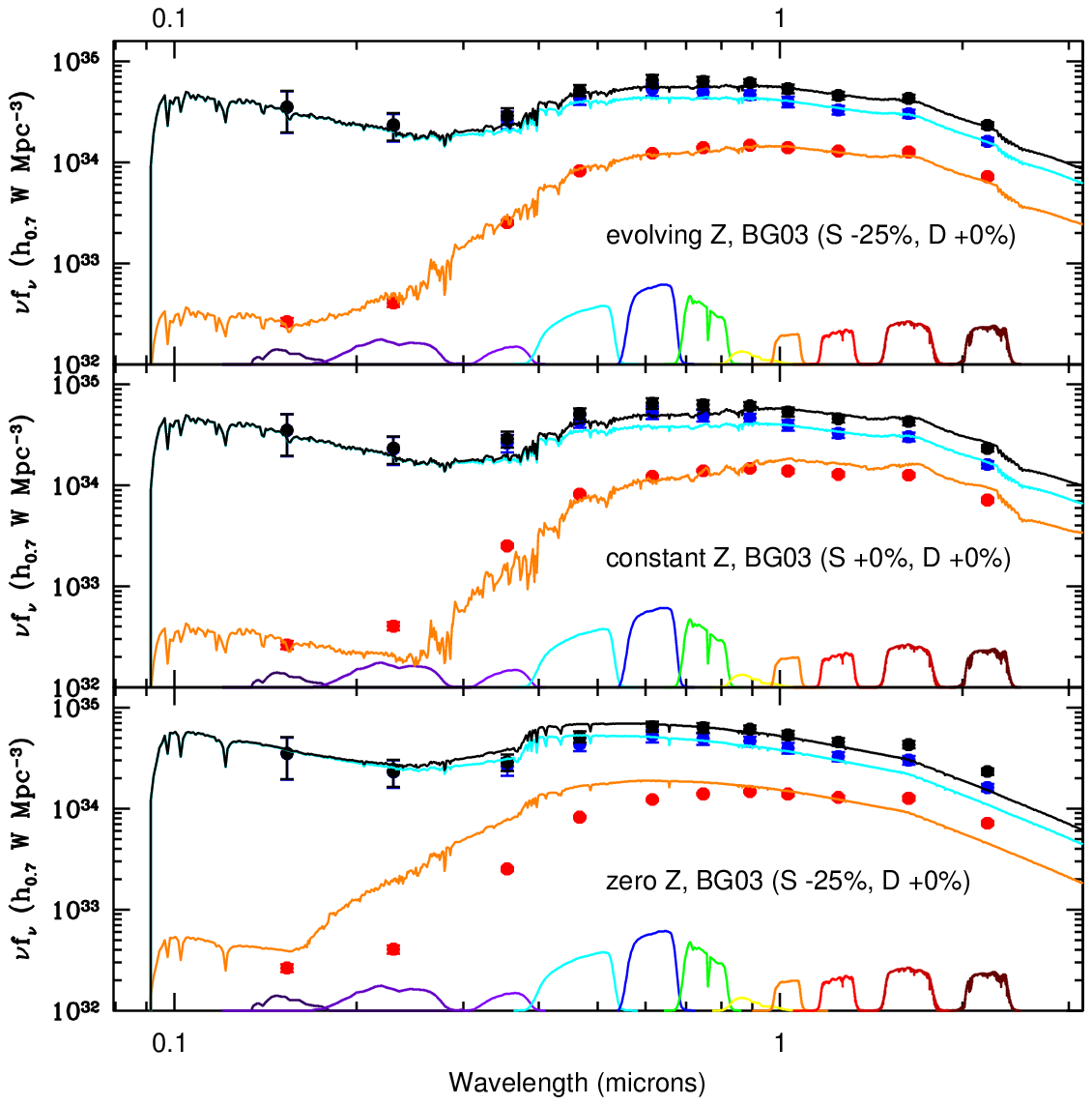,width=\textwidth}}

\caption{\label{fig:model1} The zero-parameter output, assuming a BG03
  IMF for various metallicity histories as indicated and adopting the
  star-formation histories given by Eqns.~1 \& 2. Note that the
  star-formation rates have be multiplied by a factor of $0.55$ to
  convert from a Salpeter (1955) IMF to that of BG03. The data points are
  transcribed directly from the CSED reported in Driver et al.(~2012)
  where the red points represent spheroid-dominated, the blue
  disc-dominated and the black the sum of the two. The model lines are
  for spheroids (orange), discs (cyan) and the sum (black).}
\end{figure*}

\subsection{Metal/chemical enrichment history}
Perhaps the most uncertain of the above list is the appropriate
metallicity history to adopt. Here we have been guided by the
    study of Tremonti et al.~(2004) to define our gas-phase
    metallicity at redshift zero to be $Z=0.030$ and $Z=0.010$ for the
    spheroids and discs. These values were determined by noting the
    metallicity at $10^{11}$M$_{\odot}$ (predominantly spheroids) and
    at $10^{9}$M$_{\odot}$ (predominantly discs). To convert the
    given $12+\log_{10}(O/H)$ values to those shown on
    Fig.\ref{fig:metal} we adopt a solar metallicity of
    $Z_{\odot}=0.019$ with $12+\log_{10}(O/H)_{\odot}=8.9$. We then
argue that in the absence of other factors the mean metallicity will
rise approximately linearly with the cumulative cosmic star-formation
history normalised to the present day values. This ignores the
prospect of either pre-enrichment via, for example, Population III
stars, or any lag between the star-formation and the increase in
metallicity (i.e., instantaneous enrichment).

Conceptually these are loosely consistent with a closed-box model for
spheroids (i.e., rising to a metallicity close to typical yields), and
an infall model for discs or one in which the disc is gradually
growing from a large gas ``reservoir'' (e.g., perhaps analogous to the
``equilibrium model'' put forward by Dav{\'e}, Finlator \& Oppenheimer
2012). To explore the bounds and importance of this metal enrichment,
however, we also show the CSED predictions using our simple evolving
metallicity history and for constant metallicity at the highest and
lowest values. Fig.~\ref{fig:metal} shows the implied metallicity
histories derived from Eqns.~1\&2, for the two populations (as
indicated by the red and cyan solid lines). Note that the grey
    lines on Fig.~\ref{fig:metal} show the combined metallicity of all
    stars formed (grey dotted line) and the integrated gas-phase
    metallicity (grey solid line). This shows interesting behaviour in
    that the average gas-phase metallicity (of the gas about to form
    stars) peaked at $z \sim 2$.

Constraints from the literature on the mean metallicity at
intermediate and high redshift are minimal, however we note that
Zahid, Kewley \& Bresolin (2011) find from a sample of 1350 galaxies
drawn from DEEP2 that massive systems have comparable gas-phase metallicity at $z=0.8$ to local systems, while low
mass systems have a gas-phase metallicity reduced by
  0.15dex. However Erb et al.,~(2006) find that the implied
      gas-phase metallicity, for massive, i.e.,
      spheroidal-like systems, at $z \sim 2$ is approximately half
  that at $z=0$. Both of these results are crudely consistent with our
  inferred metallicity history if one equates (as we explicitly do),
  the massive systems to spheroids and low mass systems to discs.
  Note that one natural byproduct of this is that as
      intermediate mass systems have both a spheroid and a disc
      component their systemic metallicity will lie somewhere between
      the two extremes and exhibit strong radial gradients as one
      moves from the central spheroid component to the outer disc
      component. As the mean bulge-to-total ratio increases fairly
      smoothly with stellar mass this naturally gives rise to the
      mass-metallicity relation (Tremonti et al.2004). Note that the
  $\pm 10$\% error ranges shown on Fig.~\ref{fig:metal} are purely
  indicative as the actual ranges are poorly constrained.

\subsection{Stellar population synthesis}
To construct the redshift zero CSED the {\sc pegase.2} code (Fioc \&
Rocca-Volmerange~1997; 1999) was used to produce a series of
single-stellar population (SSP) templates with an appropriate range of
metallicities (Z = 0.000 to 0.025 in 0.001 intervals) and with the {\sc
  pegase.2} default steps in ages, (i.e., roughly logarithmic from
0-20Gyrs).  For all SSP templates the star formation is set to a short
continuous burst over a 1Myr period with constant metallicity (leading
to the formation of $2.0\times 10^{-3}$ M$_{\odot}$ in {\sc
  pegase.2}-{\it normalised} stellar mass units). These SSP spectra
were then combined to create a library of 1\,Gyr time averaged spectra
from 0-1\,Gyr to 13-14\,Gyr in 1\,Gyr intervals, and for each
metallicity class. Note that the 0-1\,Gyr bin which dominates the FUV
and NUV region is extremely hard to model correctly because of the
rapidly changing UV flux and requires more care. Here we take the
rather simplistic approach of combining all the spectra provided by
{\sc pegase.2} in the 0-1\,Gyr range in the following manner i.e.,

~

\noindent
$0-1 {\rm Gyr}=\frac{\frac{\frac{1+2+3+...+10}{10}+20+30+...+100}{10}+200+300+...+1000}{10}$\,Myr

~

To create the CSED at any redshift we then sum all previously formed
populations, aged appropriately, drawn from the appropriate
metallicity class, and scaled by the required star-formation rate. The
modelling approach we adopt is therefore relatively simplistic and
effectively assumes all values (star-formation rate, metallicity etc)
are held constant over a 1\,Gyr time period. At this stage we feel
this is sufficient time resolution given the inherent uncertainties in
the initial assumptions (i.e., the input CSFH and AGN activity data).
CSEDs were then derived at all 13 time steps and combined to produce
simple evolution movies available from:

~

\noindent
http://www.simondriver.org/model1.gif --- evolving metallicity

~

\noindent
http://www.simondriver.org/model2.gif --- constant high metallicity

~

\noindent
http://www.simondriver.org/model3.gif --- constant zero metallicity

~

\subsection{Normalisation of the CSEDs}
In order to determine the correct normalisation we need to multiply
the output {\sc pegase.2} SSP spectra which are in units of
erg\,s$^{-1}$\,$\AA^{-1}$ by $\frac{10^9 \lambda}{10^7 0.002}$. Here
the factor $10^9$ scales to 1\,Gyr bins, the factor $10^7$ converts
erg\,s$^{-1}$ to W, the wavelength is in Angstroms, and the factor
0.002 scales the spectra to 1 solar mass. In applying Eqns.~1-3 we set
$\xi$ to 0.7 to correct the CSFHs to the BG03 IMF.

\begin{figure}

\centerline{\psfig{file=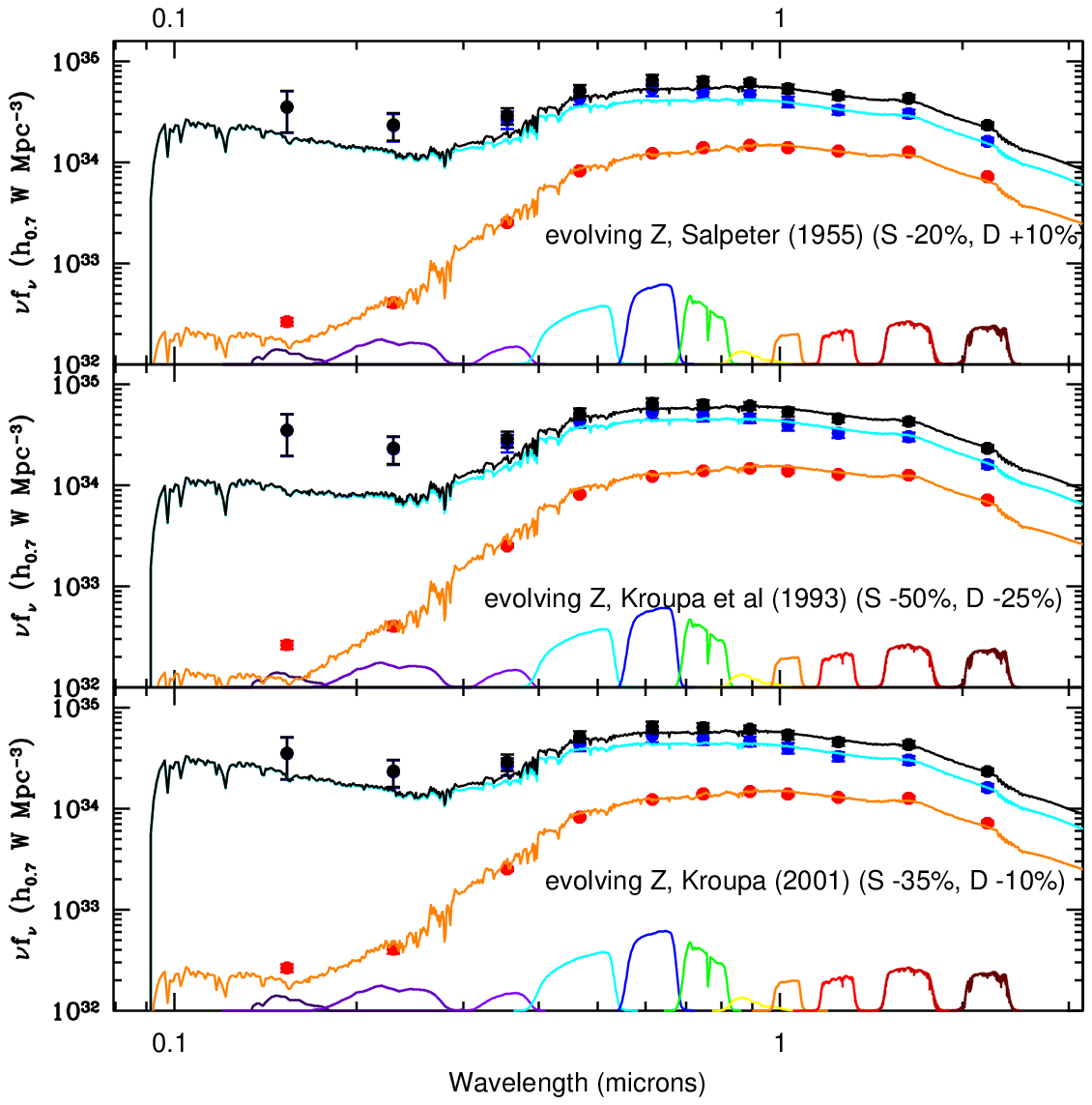,width=\columnwidth}}

\caption{\label{fig:model2} The zero-parameter output for alternative
  IMFs using the evolving metallicity shown in Fig.\ref{fig:metal} and
  adopting the star-formation histories given by Eqns.~1 \& 2. Note
  that in generating the models we modify the input star-formation by
  factors of $\times 1.0$, $\times 1.3$ and $\times 0.7$
  respectively.}

\end{figure}

Finally we allow the normalisations to float by $\pm 25$\% to account
for the uncertainty highlighted in Fig.~\ref{fig:csfh} by the grey
shading, along with uncertainties in the multiplication factors in
Table~\ref{tab:mult}, the cosmic variance in the GAMA CSED data, and
the impact of metallicity on star-formation rates. These 1
values are shown in brackets in Figs.~\ref{fig:model1} \&
\ref{fig:model2}.

\begin{figure}

\centerline{\psfig{file=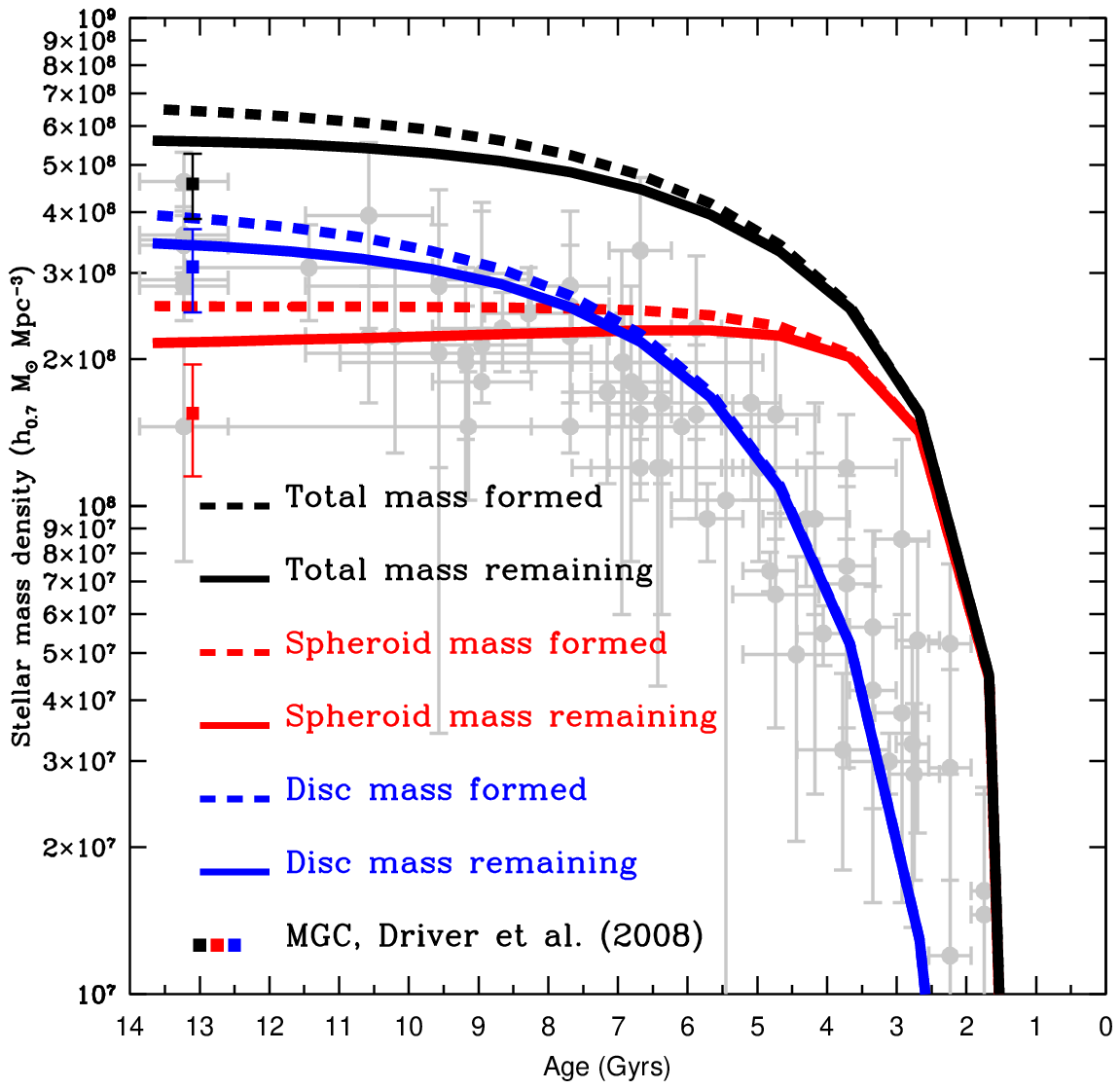,width=\columnwidth}}

\caption{\label{fig:mass} The implied build-up of stellar mass in
  spheroids and discs versus recent measurements from the Millennium
  Galaxy Catalogue. Shown in grey are the compendium of data from
  Wilkins, Trentham \& Hopkins.~2008a.}
\end{figure}

\section{Models v data}

\subsection{The CSED and adopted metallicity}
Fig.~\ref{fig:model1} shows the direct comparison of our $z=0$ CSED
models against the recent GAMA data (Driver et al.~2012), for the
three assumed metallicity histories (as indicated). The top panel,
which adopts the evolving metallicity, shows a remarkable agreement
across the full wavelength range and for both the spheroid and disc
systems. Note that in achieving these fits the spheroid data have been
renormalised downwards by 25\% which is within the specified range of
uncertainty. The central and lower panels of Fig.~\ref{fig:model1}
show the same models except for a constant high or low metallicity.
This has a negligible impact on the disc-CSED, suggesting very
    little dependency on the assumed metallicity evolution for discs
    (perhaps in part due to the low value adopted). Conversely the
impact on the spheroid CSED is quite marked with the CSED tilting
either redward or blueward for constant high or constant low
metallicity respectively. This perhaps lends some argument against any
very strong pre-enrichment phase as intermediate and low metallicity
stars in spheroids are required to produce a plausible CSED. The
obvious caveat is whether the shape of the currently adopted spheroid
CSED is significantly modified/contaminated by young disc light.

\subsection{Dependency on assumed IMF}
We now briefly explore the impact of the adopted
IMF. Fig.~\ref{fig:model2} shows the CSED predictions for the evolving
metallicity scenario using either a (top) Salpeter~(1955), (centre)
Kroupa et al.~(1993), (bottom) or Kroupa~(2001) IMF. Note that the
Kroupa~(2001) IMF is extremely close in form to the Chabrier~(2003)
IMF and therefore the Kroupa (2001) prediction can be taken as
representative of both.

Essentially all IMFs provide an equally good fit to the CSED except in
the UV regime. At wavelengths longer than $u$-band ($>0.4\mu$m) the
resulting shape is not particularly sensitive to the detailed shape of
the IMF. This is mainly because at $z=0$ stars close to solar
luminosity, where the IMF is least contentious, are dominating most of
the CSED.  Systems at the very high-mass end which formed at high
redshift will no longer be contributing to the CSED whereas very
low-mass stars are yet to dominate the near-IR flux. The CSED is
therefore unable to constrain the IMF other than the normalisations
required for these IMFs are generally higher than for the models based
on BG03.

\subsection{Stellar mass history}
Fig.~\ref{fig:mass} shows the implied build-up of stellar mass in
spheroids and discs (and combined), as indicated. Note that we show
both the total cumulative stellar mass formed (dashed lines), along
with that remaining based on default {\sc pegase.2} assumptions as to
mass-loss (solid lines). Also shown are the direct empirical stellar
mass measurements from the Millennium Galaxy Catalogue (Liske et
al.~2003; Driver et al.~2005), which includes corrections for dust
attenuation (Driver et al.~2008). The agreement is reasonable with the
discs agreeing with the MGC data to within the error and the spheroid
mass over-predicting the MGC value by a modest amount.  It is worth
noting from Fig.~\ref{fig:mass} that the stellar mass of spheroids is
actually declining, with mass-loss having exceeding mass-gain for the past 9
billion years. For the discs, the two almost exactly balance, such
that the overall stellar mass density appears to asymptote to a
constant value around the present day. 

Also shown in Fig.~\ref{fig:mass} as grey shaded data are the
compendium of total stellar mass estimates given in Wilkins, Trentham
\& Hopkins (2008a). These data clearly fall significantly below the
black shaded line, highlighting a significant discrepancy between the
total stellar mass inferred from the cosmic star-formation history and
that derived from direct empirical constraints. This offset is well
known and discussed in detailed in Wilkins, Trentham \& Hopkins
(2008a), here we make two additional comments: (1) the shape of the
data and the black curve do broadly agree with a $\times 2$ offset at
almost all ages, (2) the $z=0$ data from the MGC includes detailed
dust corrections for both the optically thin {\it and} optically thick
regions and is typically a factor of $\times 2$ higher than most local
measurements. It is possible then that the values from the literature
are missing mass embedded in optically thick regions. Perhaps a more
likely explanation, also put forward by Wilkins et al.~(2008) is that
the IMF was simply lighter at earlier times. This would reconcile
quite nicely as the low-z CSED is fairly impervious to the very low
mass-end of the IMF.

\subsection{Discussion}
At this point we have a simple heuristic model which adopts two simple
axioms motivated by the physically distinct appearance of spheroids
and discs in the nearby Universe. These axioms combined with the
empirical compendium of AGN and cosmic star-formation activity/history
are able to reproduce the $z=0$ CSEDs of spheroids and predict the
mean mass and metallicity evolution of present day discs and
spheroids. The model also provides a complete description of the
energy output from stars within those systems which will eventually
make up the local spheroid population (projenitors) as a function of
redshift, the metallicity build-up, and suggests key cross-over epoch
at $z \sim 1.6$ between the hot and cold mode evolution. This later
transition redshift is consistent with the obvious change in
morphologies seen in HST images at this redshift (e.g., Driver et al.,
1998, figure 3).

However an obvious weakness is that the model provides no clear
prediction of the morphological, size- and shape- evolution, merger
rates, or the clustering of the galaxy population. Furthermore the
model does not stipulate the actual mechanism by which star-formation
is occurring and for the hot mode could be some combination of
monolithic collapse, major merging, and/or clump migration. Within the
recent literature the exact status of the $z \sim 2$ population is
also unclear. Deep IFU studies (e.g., Forster Schreiber et
al.,~2009,2011) find that the majority of star-formation at $z \sim 2$
appears to be taking place in rotating clumpy disc structures with no
obvious central bulge component. Similarly Chevance et al.,~(2012, see
also Weinzirl et al.,~2011) from a study of 31 high-z galaxies, find
that the S\'ersic indices are significantly flatter than one would
expect for low-z spheroids and obvious disc structures are present in
many cases. From our Fig.~1 we can see that at $z\sim 2$ we are still
within the epoch where spheroid formation should be
dominating. However spheroid formation does not necessarily imply
spheroid morphologies until after some unspecified time-lag in which
the system settles. In fact violently star-forming systems will
inevitably appear blue, asymmetrical, gas-rich, and dusty, i.e., quite
disc-like in many aspects. Other studies, e.g., van Dokkum (2008),
find that 45\% of massive galaxies at $z\sim 2.3$ do indeed have
evolved stellar populations, little or no ongoing star-formation, and
compact early-type morphologies. Hence a picture of a spheroid
population emerging from a highly turbulent progenitor phase around $z
\sim 2$ appears to be qualitatively consistent with our
model. Alternatively our model may need to be adjusted to allow for
gas infall and disc-formation from the outset with some fraction of
the disc-formed stars merging into bulges, i.e., a relaxation of the
maximal spheroid formation axiom. This would have the net effected of
also increasing the cross-over redshift to $>1.6$

A further intriguing observations is that high-z spheroids are
significantly more compact than nearby ellipticals by factors of
$\times 3-4$ at fixed stellar mass (e.g., Daddi et al.,~2005; van
Dokkum et al.,~2008 etc). Within our scenario this could be consistent
with the high-z sample being ``naked''-bulges yet to grow discs or yet
to be ``puffed-up'' through successive minor merger interactions or
adiabatic expansion. These two pathways, disc-growth verses
``puffing'', are likely to be strongly environmentally dependent with
minor mergers more frequent in high density environments, and gas
infall more prevalent in low-density environments. A particularly
interesting comparison might therefore be the mass-size relation of
high-z spheroids to low-z bulges. 

With the caveat that the morphology and size evolution within our
model is unspecified we nevertheless appear to have a prediction of
the energy output of spheroid and discs projenitors over all epochs
(Fig.~\ref{fig:model1} top panel), the mean gas-phase
  metallicity history for each population (Fig.~\ref{fig:metal}), and
  the build-up of stellar mass (Fig.~\ref{fig:mass}). Whether one can
  readily distinguish these populations observational however is an
  open question.

Finally it is worthwhile reiterating that this model contains {\it no
  tunable parameters nor any dependency on initial conditions beyond
  the underlying cosmology}. The model is built entirely from
empirical data and provides a fully consistent empirical scaffolding
upon which more physically motivated models can be built. Our
conclusion is that the initial axioms on which the model is based are
viable and the star-formation histories defined are tenable.

Further studies of the variation of the $z=0$ CSED and its dependency
on environment should enable an investigation into dependencies on
clustering, and to assess whether star-formation proceeds more rapidly
or whether it is merely the relative mix of spheroid v disc formation
which is changing. Similarly, using observations at intermediate
redshift, it should be possible to compare data from high-z studies to
the predictions of our two-phase model. Both of these avenues will be
explored in future papers.

\begin{figure}

\centerline{\psfig{file=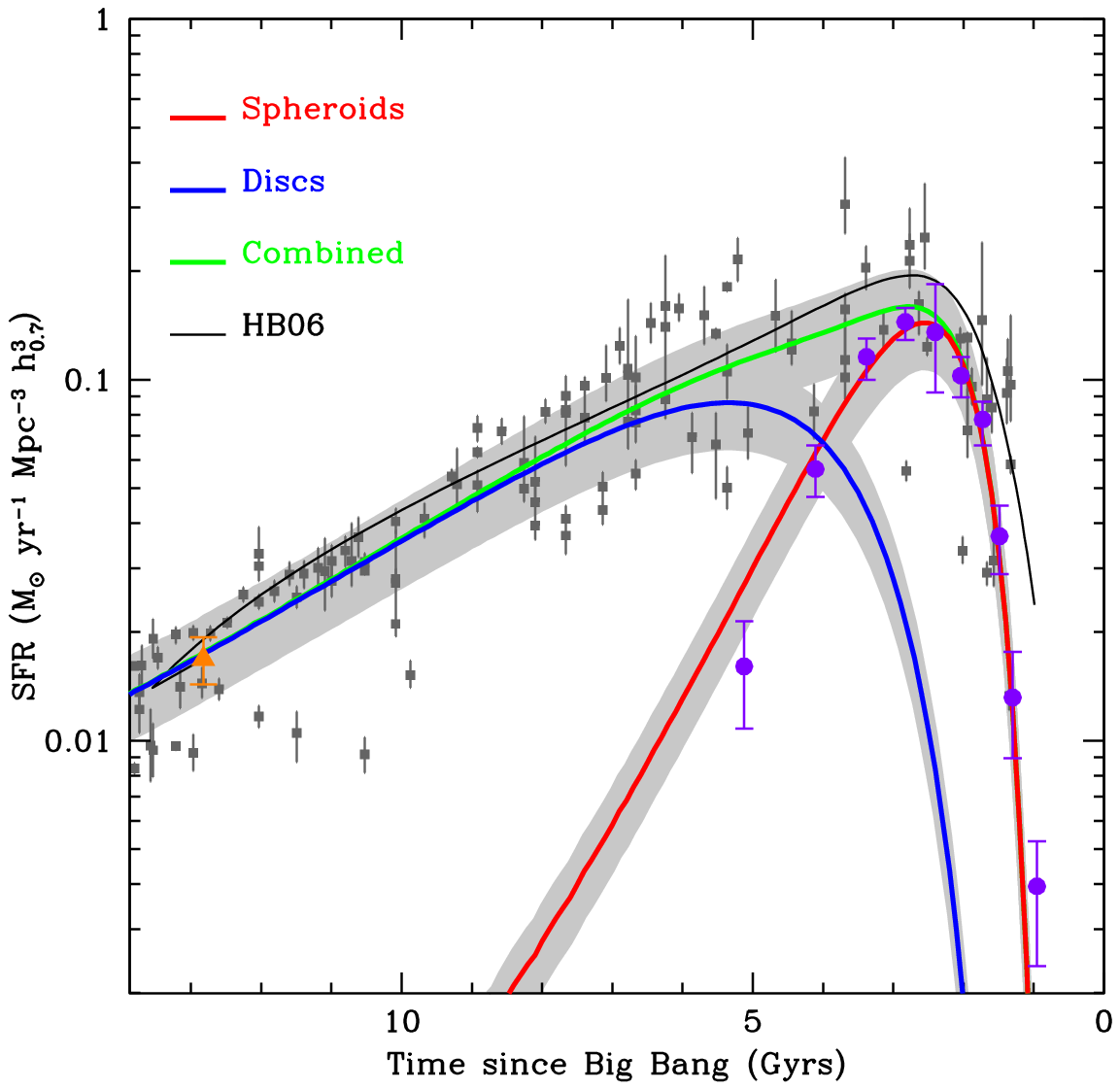,width=\columnwidth}}

\caption{As for Fig.~\ref{fig:csfh} (lower) except with the spheroid
  star-formation history down-weighted by 25\% by incorporating the
  CSED constraints from Fig.~\ref{fig:model1}
  (upper). \label{fig:csfhadj}}
\end{figure}

\section{Conclusions}
From two very simple axioms: (1) that AGN activity traces spheroid
formation, and (2) that the CSFH is dominated by spheroid formation at
high redshift, we are able to derive simple expressions for the cosmic
star-formation histories of spheroids and discs. Following comparisons
to the $z=0$ CSED for spheroids and discs we find a modest downward
adjustment of 25\% provides the optimal fit resulting in our final recommended star-formation histories of:
\begin{eqnarray}
\dot{\rho}_{S}=\xi 0.77 \times 10^{-5} h_{0.7}^{3} (\frac{21.86}{t_{\rm Gyrs} h_{0.7}})^{8.57}\exp(-\frac{21.86}{t_{\rm Gyrs} h_{0.7}}) \\
\dot{\rho}_{D}=\xi 1.80 \times 10^{-3} h_{0.7}^{3}(\frac{29.39}{t_{\rm Gyrs} h_{0.7}})^{5.50}\exp(-\frac{29.39}{t_{\rm Gyrs} h_{0.7}})
\end{eqnarray}
where $\xi$ is the IMF multiplier as given in Table~\ref{tab:mult}.
Fig.~\ref{fig:csfhadj} shows these final relations compared to the
compendium of data provided by Hopkins \& Beacom (2006) and despite
the renomalisation of the spheroid star-formation history still
provide a perfectly satisfactory description of the global CSFH.

Adopting a Baldry \& Glazebrook~(2003) IMF and using these
expressions to predict the $z=0$ CSED, we are able to provide a
satisfactory explanation of the observed CSEDs of spheroids and discs
from the FUV to the $K$-band.

The corollary of the simplicity of the two-phase model, however, is
that it lacks any prediction of the clustering signature,
environmental dependencies, or the merger histories, although these can
be built in at a later stage. Perhaps the key gain, in an era of
hidden tunable parameters, is that with the adoption of a universal
IMF and a stellar evolution code there are essentially no free
parameters. Strictly speaking this is not precisely true as the
detailed modelling of stellar evolution typically comes with options
and there is arguably a choice of IMFs and also whether it is
Universal or varies over cosmological time (see for example Wilkins et
al.~2008a, 2008b; Gunawardhana et al.~2011). On this last subject of
the IMF it is worth reiterating that longward of $0.4\mu$m the $z=0$
CSED is not sensitive to the high-mass shape of the IMF
(unless taken to the extreme). This is because at almost all wavelengths,
in the declining star-formation era today, the CSED is dominated by either
the tip of the main sequence, which lies just below a solar mass and well
above the mass range of contention (mid-optical to NIR), or the most
recently formed stars (FUV to mid-optical).

As a byproduct, the two-phase model also provides the CSED of
spheroids and discs at every epoch in the Universe, along with the
prediction of a clear-cut transition redshift at around $z \approx
1.7$ where galaxy evolution switches from evolution being dominated by
major mergers to evolution being dominated by cold gas infall. Future
work will include a broader wavelength baseline, bulge-disc
decompositions, inclusion of the AGB energy output, and development of
the model via comparisons to selected external data.

\section*{Acknowledgments} 
We thank the referee for insightful comments during the refereeing
process which has helped improve the paper.

\section*{References}

\reference Allen P.D., Driver S.P., Graham A.W., Cameron E., Liske J., de Propris R., 2006, MNRAS, 371, 2

\reference Abadi M.G., Navarro J.F., Steinmetz M., Eke V.R., 2003, ApJ, 591, 499

\reference Agertz O., Romain T., Moore B., 2009, MNRAS, 397, 64

\reference Baldry I.K., Glazebrook K., ApJ, 2003, 593, 258

\reference Baldry I.K., 2004, ApJ, 600, 681

\reference Barnes J.E., Hernquist L., 1996, ARA\&A, 30, 705

\reference Behroozi P.S., Wechsler R.H., Conroy C., 2012, ApJ, submitted (arXiv:1207.6105)

\reference Brown M.J.I., Dey A., Jannuzi B.T., Brand K., Benson A.J.,
Brodwin M., Croton D., Eisenhardt P.R., 2007, ApJ, 654, 858

\reference Brown M.J.I., et al., 2008, ApJ, 682, 937

\reference Bundy K., Fukugita M., Ellis R.S., Kodama T., Conselice C., 2004, ApJ, 601, 123

\reference Chabrier G., 2003, PASP, 115, 763

\reference Chevance M., et al.,~2012, ApJ, 754, 24

\reference Cole S. et al., 2001, MNRAS, 326, 255

\reference Conselice C.J., Blackburne J.A., Popovich C., 2005, ApJ, 620, 564 

\reference Cook M., Lapi A., Granato G.L., 2009, MNRAS, 397, 534

\reference Cook M., Barausse E., Evoli C., Lapi A., Granato G.L., 2010b, MNRAS, 402, 2113

\reference Cook M., Evoli C., Barausse E., Granato G.L., Lapi A., 2010a, MNRAS, 402, 941

\reference Daddi E., et al., 2005, ApJ, 626, 680

\reference Dav\'e R., Finlator K., Oppenheimer B.D., 2012, MNRAS, 421, 98

\reference Dekel A., et al., 2009, Nature, 457, 451

\reference De Propris R., Liske J., Driver S.P., Allen P.D., Cross N.J.G., 2005, MNRAS, AJ, 130, 1516

\reference De Propris R., Conselice C., Liske J., Driver S.P., Patton D.R., Graham A.W., Allen P.D., 2007, ApJ, 666, 212
 
\reference De Propris R., et al., 2010, AJ, 139, 794

\reference Domenech-Moral M. Martinez-Serrano F.J. Dominguez-Tenreiro R., Serna A., 2012, MNRAS, 412, 2510

\reference Driver S.P. Fernandze-Soto A., Couch W.J., Odewahn S.C.,
Windhorst R.A., Phillipps S., Lanzetta K., Yahil A., 1998, ApJ, 496,
93

\reference Driver S.P., 1999, ApJ, 526, 69

\reference Driver S.P., Liske J., Cross N.J.G., De Propris R., Allen P.D., 2005, MNRAS, 360, 81

\reference Driver S.P., et al., 2006, MNRAS, 368, 414

\reference Driver S.P., Popescu C.C., Tuffs R.J., Liske J., Graham A.W., Allen P.D., de Propris R., 2007, MNRAS, 379, 1022

\reference Driver S.P., Popescu C.C., Tuffs R.J., Graham A.W., Liske J., Baldry I., 2008, ApJ, 678, 101

\reference Driver S.P., Robotham A.S.G., 2010, MNRAS, 407, 2131

\reference Driver S.P., et al., 2012, MNRAS, in press (astro-ph/1209.0259)

\reference Driver S.P., et al., 2011, MNRAS, 413, 971

\reference Elmegreen B.G.,  Bournaud F., Elmegreen D.M., 2007, ApJ, 658, 67

\reference Erb, D., Shapley A., Pettini M., Steidel C.C., Reddy N.A., Adelberger K.L., 2006, ApJ, 644, 813

\reference Ferrarese L., Ford H., 2005, SSRv, 116, 523

\reference Fioc M., Rocca-Volmerange B., 1997, A\&A, 326, 950

\reference Fioc M., Rocca-Volmerange B., 1999, (arXiv:9912179) 

\reference Forster Schreiber N.M., et al.~2011, AJ, 739, 45

\reference Forster Schreiber N.M., et al.~2009, ApJ, 706, 1364

\reference Gadotti D., 2009, MNRAS, 393, 1531


\reference Geller M.J., Diafero A., Kutz M.J., Dell'Antonio I.P., Fabricant D.G., 2012, AJ, 143, 102  

\reference Graham A.W., 2011, in Planets, Stars and Stellar Systems, (Publ: Springer)

\reference Governato F., et al., 2010, Nature, 463, 203

\reference Governato F., et al., 2012, MNRAS, 422, 1231

\reference Hill, D., et al., 2011, MNRAS, 412, 765

\reference House E.L., et al., 2011, MNRAS, 415, 2652

\reference Hopkins A.M., Beacom J.F., 2006, ApJ, 651, 142

\reference Hopkins P.F., Hernquist L., Cox T.J., Di Matteo T., Robertsan B., Springel V.,  2006, ApJS, 163, 1

\reference Hopkins P.F., Hernquist L., Cox T.J., Keres D., 2008a, ApJS, 175, 356

\reference Hopkins A.M., McClure-Griffiths N.M., Gaensler B.M., 2008b, ApJ, 682, L13

\reference Hopkins P.F., Cox T.J., Younger J.D., Hernquist L., 2009, ApJ, 691, 1168 

\reference Hubble E., 1926, ApJ, 64, 321

\reference Hubble E., 1936 in Realm of the Nebulae (Pub: Yale Uni. Press)

\reference Keres D., Katz N., Weinberg D.H., Dave R., 2005, MNRAS, 363, 2

\reference Kistler M.D., Yuksel H., Beacom J.F., Hopkins A.M., Wyithe J.S.B., 2009, ApJ, 705, 104

\reference Koda J., Milosavljevic M., Shapiro P.R.,. 2009, ApJ, 696, 254

\reference Komatsu et al., 2011, ApJS, 192, 18

\reference Kroupa P., 1993, MNRAS, 262, 545

\reference Kroupa P., 2001, MNRAS, 322, 231

\reference Larson R.B., 1976, MNRAS, 176, 31

\reference L'Huillier B., Combes F., Semelin, B, 2012, A\&A, 544, 68

\reference Liske J., Lemon D.J., Driver S.P., Cross N.J.G., Couch W.J., 2003, MNRAS, 344, 307

\reference Lackner C.N., Gunn J.E., 2012, MNRAS, 421, 2277

\reference Martig M., Bournaud F., 2010, ApJ, 714, 275

\reference Navarro J.F., Steinmetz M., 2000, ApJ, 538, 477


\reference Patton D.N., et al., 2002, ApJ, 565, 208

\reference Pereira, E.S., Miranda O.D., 2011, MNRAS, 418, 30

\reference Polletta M., et al., 2006, ApJ, 642, 673

\reference Popescu C.C., Tuffs R.J., Dopita M.A., Fischera J., Kylafis N.D., Madore B.F., 2011, A\&A, 527, 109

\reference Rafferty D.A., Brandt W.N., Alexander D.M., Xue Y.Q., Bauer F.E., Lehmer B.D., Luo B., Papovich C., 2011, ApJ, 742, 3

\reference Ravindranath S., et al., 2006, ApJ, 652, 963

\reference Richards G., et al 2006, AJ, 131, 2766

\reference Robotham A.S.G., Driver S.P., 2011, MNRAS, 413, 2570

\reference Salpeter E.E., 1955, ApJ, 121, 161

\reference Sancisi R., Fraternall F., Oosterloo T., van der Hulst J.M., 2008, A\&A Rv, 15, 189 

\reference Scannapieco C, White S.D.M., Springer V., Tissera P.B., 2011, MNRAS, 417, 154

\reference Simard L., Mendel J.T., Patton D.R., Ellison S.L., McConnachie A.W., 2011, ApJS, 196, 11 

\reference Strateva I., et al., 2001, AJ, 122 1861

\reference Tasca L.A.M., White S.D., 2011, A\&A, 530, 106 

\reference Tinsley B.M., Larson R.B., 1978, ApJ, 221, 554

\reference Tremonti C.A., et al., 2004, ApJ, 613, 898

\reference Triester E, Urry M.C., Schawinski K., Cardamone C.N., Sanders D.B., 2010, ApJ, 722, 238

\reference Tuffs R., Popescu C.C., Volk HJ., Kylafis N.D., Dopita M.A., 2004, A\&A, 419, 821

\reference van den Bergh S., 2002, PASP, 114, 797

\reference van Dokkum P., et al.,~2008, ApJ, 677, 5

\reference Vika, M., Driver S.P., Cameron E., Kelvin L., Robotham A.S.G., 2012, MNRAS, 419, 2264

\reference Weinzirl T., et al., 2011, ApJ, 743, 87

\reference White S.D.M., Navarro J.F., 1993, Nature, 366, 429

\reference Wilkins S., Trentham N., Hopkins A.M., 2008a, MNRAS, 385, 687

\reference Wilkins S., Hopkins A.M., Trentham N., Tojeiro R., 2008b, MNRAS, 391, 363 

\reference Yuksel H., Kistler M.D., Beacom J.F., Hopkins A.M., 2008, ApJ, 683, 5

\reference Zahid H.J., Kewley L.J., Bresolin F., 2011, ApJ, 730, 137

\reference Zolotov A., et al., 2012, ApJ, submitted (astro-ph/1207.0007)

\reference Zwicky F., 1957, in Morphological Astronomy (Pub: Springer)

\end{document}